\documentstyle[aps,twocolumn,epsf]{revtex}

\renewcommand{\vec}{\bbox}

\def\beqra{\begin{eqnarray}}
\def\eeqra{\end{eqnarray}}
\def\beq{\begin{equation}}
\def\eeq{\end{equation}}
\def\ds{\displaystyle}

\def\L{\Lambda}

\def \lta {\mathrel{\vcenter
     {\hbox{$<$}\nointerlineskip\hbox{$\sim$}}}}
\def \gta {\mathrel{\vcenter
     {\hbox{$>$}\nointerlineskip\hbox{$\sim$}}}}

{

\def\half{\mbox{\small $\frac{1}{2}$}}
\sloppy

\begin{document}
\title{Critical Slowing Down from $T\neq 0$ Wilson RG}

\author{Massimo Pietroni}

\address{INFN - Sezione di Padova,
Via F. Marzolo 8, I-35131 Padova, Italy}

\date{\today}

\maketitle

\begin{abstract}
The Thermal Renormalization Group can be employed to study the dynamics
of $T\neq 0$ Quantum Field Theories close to second order 
phase transitions, where neither resummed perturbation theory nor
first principle lattice simulations can be employed.  

As an example, I discuss the computation of  the plasmon damping rate in 
the scalar field theory from $T\gg T_C$ down to $T_C$.  
As the critical point is approached, the lifetime of  long wavelength 
thermal fluctuations diverges.

Taking this effect into account, the notion of quasiparticle, and kinetic
approaches,  can be extended close to the critical regime. 
The consequences on the scenarios for topological defect formation in the
early universe are also discussed.
\end{abstract}

\narrowtext


\section{Critical speeding up vs. critical slowing down}
\label{sec1}
The last few years have seen a dramatic progress in the understanding
of the {\it static} properties of relativistic quantum field theory (QFT)
at high 
temperature. Perturbative computations, lattice simulations, and the 
Wilson Renormalization Group, have been successfully applied to many 
different problems, {\it e.g.} to the investigation of the free energy in 
the scalar theory, in the electroweak standard model, and in QCD.
On the other hand, there are many interesting {\it dynamical} phenomena, 
like the generation of the cosmological baryon asymmetry or the formation of 
topological defects in a second order phase transition, in which time plays a 
crucial role and a real time formulation (either in or out of equilibrium) is 
mandatory. Since first principle lattice simulations 
in the real time formalism are not feasible and perturbation theory is
plagued by severe infrared (IR) problems, semi-phenomenological approaches, 
mainly based on Langevin-type equations, have been employed so far.

The Thermal Renormalization Group (TRG) introduced in \cite{DP1} offers a 
promising alternative to this strategy. First of all, it is formulated in the 
real-time formalism, where non-static quantities are more conveniently
computed. Second, IR divergences are kept under control by the introduction 
of an explicit momentum cut-off. Third, by lowering the cut-off, the IR regime
is smoothly approached via a continuos sequence of effective theories
describing the dynamics of the soft modes only, the hard ones having been
integrated out. This makes the TRG particularly suited to the investigation
of those dynamical systems which are dominated by the soft modes.

In this talk I will discuss the computation of the plasmon damping rate in 
the real scalar theory $\lambda \phi^4/4!$, from temperatures $T \gg T_C$ down
to the second order phase transition for $T\rightarrow T_C$. The plasmon 
damping rate is defined as
\beq
\gamma_{\vec{k}} (T) \equiv \frac{\Pi_I(\omega_{\vec{k}},\vec{k})}{2 
\omega_{\vec{k}}}\;,
\label{gammadef}
\eeq
where $\omega^2_{\vec{k}}=\vec{k}^2+m(T)^2$, $m(T)$ is the thermal mass  and 
$\Pi_I$ is the imaginary part of the self-energy.

The physical interpretation of the dynamical quantity $\gamma_{\vec{k}}(T)$ 
was given by Weldon long ago \cite{W} (see also \cite{BLL}): if the plasma 
is slightly out of thermal equilibrium then $\gamma_{\vec{k}}(T)$ gives half
the relaxation rate of the quasiparticle distribution 
function  to its equilibrium value,
\beq
\frac{d \,\delta n_{\vec{k}}}{d t} = - 2 \gamma_{\vec{k}}(T) 
\delta n_{\vec{k}}\,
\eeq
where $\delta n_{\vec{k}}$ is the deviation of the distribution function from
equilibrium $\delta n_{\vec{k}} = n_{\vec{k}} - n^{eq}_{\vec{k}}$. 

In the following, I will be interested in the behavior of long wavelength
fluctuations as the critical temperature is approached, so I will concentrate 
on the damping rate for vanishing spatial momentum, $\gamma(T) \equiv
\gamma_{\vec{k}=0}(T) $.

In perturbation theory, $\gamma(T)$ is a two-loop effect, and is given by
\cite{P,HW}
\beq
\gamma_{p.t.} = 
\frac{1}{1536 \pi}\, l_{qu} \,\lambda^2 T^2\,
\label{gammacl}
\eeq
where $l_{qu}=1/m(T)=\left(\frac{\lambda T^2}{24}\right)^{-1/2} $ is 
the Compton wavelength.

In refs. \cite{ASBJ1} it was shown that the above two-loop result can
be reproduced in the 
classical theory provided that the Compton wavelength $l_{qu}$ is 
identified with  the classical correlation length $l_c$. 
This can be understood realizing that $\gamma$ probes
the theory at scales $\omega = m(T) \ll T$ 
(if $\lambda$ is perturbatively small), where the Bose-Einstein 
distribution function is approximated by its classical limit,
\beq
N(\omega) = \frac{1}{e^{\beta \omega} - 1} \rightarrow 
\frac{1}{\beta m(T)}\;\;\;\;\;\;\;\;\,\;\;({\rm if} \; \beta \omega \ll 
1)\;,
\label{class}
\eeq
with $\beta = 1/ T$. Considering for instance the 1-1 
component of the propagator in the real-time formalism,
\beq D_{11} = {\cal P} \frac{1}{k^2-m^2} - 2\pi i  \delta(k^2-m^2)
\left( \frac{1}{2} + N(|k_0|)\right),
\label{pro}
\eeq
we see that when the loop momenta are $k_0 \ll T$ the `statistical' 
contribution to the imaginary part of the propagator dominates over the 
`quantum' one, {\it i.e.} $N \gg 1/2$, and the leading order result can be 
obtained neglecting the $T=0$ quantum contributions to the loop
corrections. 

The above argument has been employed to motivate the use of 
classical equations of motion in the study of the evolution of long 
wavelength modes in scalar and gauge theories. More recently, this 
approach has been improved by many authors including the effect of 
Hard Thermal Loops in the equations of motion for the `soft' modes (see
for instance \cite{HTL,BJ2}).
The separation between hard and soft modes has been made explicit in ref.
\cite{BJ2} by introducing a cutoff $\Lambda$ such that 
$m(T) \lta \Lambda < T$. 

In the $T\gg T_C$ limit we have $m(T)/T \simeq \left(\lambda /24 \right)
^{1/2}$, therefore the `statistical' dominance is valid only up to 
$O(\lambda^{1/2})$ corrections. On the other hand the critical region 
corresponds, by definition, to $m(T) \rightarrow 0$, and the statistical limit
is exact in this case.

What does perturbation theory predict in this limit? If we trust the 
two-loop/classical expression in eq. (\ref{gammacl}), then we see that 
$\gamma$ diverges as the correlation length. Physically, this would mean that 
the lifetime of long wavelength fluctuations gets shorter and shorter 
as the critical temperature is approached.
However, it is well known that (resummed) perturbation theory cannot be
trusted close to the critical point, due to the divergence of  its effective
expansion parameter, ${\it i.e.} \; \lambda T/m(T)$.

A more reliable indication of what's going on can be obtained from the
phenomenological theory of dynamical critical phenomena \cite{LK,HH} 
developed by Landau in
the fifties. Let's consider a non-relativistic, classical scalar
theory described by an order parameter $\eta(t,\vec{r})$ and the 
free energy 
\[{\cal F}(\eta)=\int d^3 \vec{r} \left[ 
\half (\partial_i \eta)^2 + \alpha (T-T_C) \eta^2\right]\,.
\]
If $\eta$ is slightly displaced from 
the minimum of ${\cal F}$ (at $\eta =0$), then it is 
reasonable to assume that equilibrium will be restored at a rate given by
\beq
\frac{\partial \eta(t, \vec{r})}{\partial t} = - \Gamma \frac{\delta {\cal F}}{
\delta \eta(t,\vec{r})} + \xi(t,  \vec{r}),
\label{LANG}
\eeq
where $\Gamma$ is a phenomenological parameter which was assumed to be 
temperature independent in the original formulation of the theory, and 
$\xi(t,  \vec{r})$ is a white noise term. Eq. (\ref{LANG}) is usually quoted in
the literature as the time-dependent Ginzburg-Landau equation. 
Taking Fourier transform, we see that the $\vec{k}=0$ mode  vanishes
as $\exp(-t/\tau_0)$ whith the space-independent relaxation rate given by
\beq
\tau_0 = 2 \gamma(T) = 2 \alpha \Gamma (T-T_C),
\label{slowing}
\eeq
thus vanishing as $T \rightarrow T_C$, in open contrast with the divergent 
perturbative result.

Dynamical critical phenomena have
been the subject of  intense study by the condensed matter community 
in the seventies (see the reviews in \cite{HH}). 
The Ginzburg-Landau approach has been improved by renormalization group methods
coupled to $\varepsilon$ or $1/N$ expansions.
In general, the assumption that transport and kinetic coefficients such as 
$\Gamma$ in eq. (\ref{LANG}) are temperature independent turns out to be false.
In any known case, however, the combination $(T-T_C) \Gamma$ still vanishes.
Then, the  {\it critical slowing down} found 
in (\ref{slowing}) persists, even though with a different power law.

\section{What about 3+1 dimensional, relativistic, $T\neq 0$, QFT's?}

We have seen that the behavior of $\gamma$ predicted by perturbation theory 
is  exactly opposite to what is obtained by the Landau-Ginzburg approach. On 
the other hand, the latter is based on quite general and physically reasonable
assumptions, while it is well known that perturbation theory is unreliable
in the critical region.

In refs.\cite{E,TW,DP1} it was shown that the key effect which is 
missed by perturbation theory is the dramatic thermal renormalization of 
the coupling constant, which vanishes in the critical region.
This is just a consequence of triviality of $\lambda \phi^4$ theory, which 
implies that, in absence of any physical IR cut-off, the renormalized coupling
at zero external momentum vanishes, even if the ultraviolet cut-off, 
$\Lambda_{UV}$, is kept fixed. This is true even at $T=0$, where $\lambda$
vanishes as  $1/\log(m/\Lambda_{UV})$ when $m\rightarrow 0$. 
At $T\neq 0$, due to the effective three 
dimensionality of the theory in the critical region, we have a stronger, 
linear, dependence, 
$\lambda(T) \sim m(T)$.

In the framework of 
Wilson Renormalization Group this can also be understood as follows. 
The IR regime of the four-dimensional
field theory at $T=T_C$ is related to that of the three-dimensional
theory at $T=0$. In particular, the three-dimensional running coupling 
is obtained from the four-dimensional one by \cite{TW}
\[
\lambda_{3D}(\Lambda) = \lambda(\Lambda) \frac{T}{\Lambda}\,,
\]
where $\L$ is the running parameter.
At the critical point, $\lambda_{3D}$ flows in the IR to the 
Wilson-Fischer fixed point value $\lambda^*_{3D} \neq 0$, so that the 
four-dimensional coupling vanishes,
\[
 \lambda(\Lambda)\rightarrow \frac{\Lambda}{T} \lambda^*_{3D} 
\;\;\;({\rm for}\; T\simeq T_C \;{\rm and}\; \Lambda \rightarrow 0)\,.
\]
The critical exponent governing the vanishing of $\lambda$ is the same as
that for $m(T)$, so that the ratio $\lambda(T)T/m(T)$ goes to a finite
value at $T_C$, and a second order phase transition is correctly
reproduced.

We will see that the running of the coupling constant for $T\simeq T_C$
is crucial also in turning the divergent behavior of eq. (\ref{gammacl})
into a vanishing one. 
As a first rough ansatz one could just replace the tree coupling 
$\lambda$ with the thermally renormalized one in eq. (\ref{gammacl}) and 
readily 
see that the critical speeding up is indeed turned into the expected slowing 
down.
However, as we will see, this gives only a qualitatively correct answer,
due to the logarithmic singularity of the on-shell imaginary part when $m(T)$ 
vanishes. 

\section{The TRG}

Let me now present in some detail the computation of $\gamma(T)$ in the TRG 
framework.
This method allows a computation of the damping rate for any value
of $T$, from very high values, where perturbation theory works, to the 
critical region, where renormalization group methods like those of \cite{HH} 
are necessary to resum infrared divergencies. Remarkably, the TRG is 
applicable also
in the intermediate region, in which none of these methods can be employed.

Wilson's RG idea was originally formulated on the lattice \cite{Wi},
and then implemented in continuum QFT's by Polchinski \cite{Po}.
In this framework, it is based on the introduction of an explicit IR cut-off, 
via the modification
of the tree level propagator. Let me briefly recall how it works for the 
 scalar theory at $T=0$.
The starting point is the substitution 
\[ D(p)= \frac{1}{p^2+m^2} \rightarrow D_\L(p)=D(p) \Theta(p^2-\L^2)\;,
\]
(where $\Theta$ can be Heavyside's step function or a smooth cut-off) 
in the usual expression for the generating 
functional,
\[
Z_\L[J] = \int {\cal D}\phi \exp\left[-\left(\half \,\phi \cdot D_\L^{-1} \cdot
\phi + S[\phi] + J \cdot \phi\right)\right]\;.
\]
$Z_\L[J]$ generates Green functions in which only the modes with $p > \L$ have
been integrated out. Deriving with respect to $\L$ the generating 
flow equation is obtained
\beq
\L \frac{\partial\;\;}{\partial \L} Z_\L[J] = -\half \left(
\frac{\delta\;\;}{\delta J} \cdot \frac{\partial\;\;}{\partial \L} D_\L^{-1}
\cdot \frac{\delta\;\;}{\delta J} \right)  Z_\L[J]\,.
\label{flowT0}
\eeq
The exact equation (\ref{flowT0}) provides a non perturbative definition of
the $T=0$ QFT. It interpolates between the `bare' theory, defined at some
ultraviolet scale $\L=\L_{UV}$, and the renormalized one at $\L=0$, in 
which all quantum fluctuations with momenta $0\le p\le \L_{UV}$ are included.

A serious problem of this kind of approach is encountered when gauge theories 
are considered. Indeed, the momentum cut-off explicitly breaks gauge 
invariance, leading to $\L$-dependent modified Slavnov-Taylor  identities 
(ST) which
only in the $\L=0$ limit recover their usual form. Even though in principle 
one 
can work with this new form of the ST's, it is in practice quite non trivial. 
Moreover, no physical meaning can be assigned to the theory at non-zero
value of $\L$, and the interpretation of $\L$ as a physical resolution scale
is not possible at all.

When it comes to $T\neq 0$ QFT's, two possibilities are on the ground. One is
to work in the imaginary time formalism and  straightforwardly  apply the
same steps outlined above, getting flow equations which interpolate between
the $T=0$ bare theory and the $T\neq 0$ equilibrium QFT \cite{TW,LAS}.
As long as one is interested in static quantities in the critical regime of
non-gauge theories, this approach works very well.

The other possibility is the  TRG \cite{DP1}. The TRG flow has three unique 
characteristics:
{\it i)} it is formulated in the real time; {\it ii)} it interpolates
between the $T=0$ {\it renormalized} QFT and the $T\neq 0$ QFT in thermal 
equilibrium; {\it iii)} it is explicitly gauge invariant \cite{DP2}.
All this is achieved by exploiting the fact that the tree level propagator in 
the real-time formalism (eq. (\ref{pro})) is made up by the sum of two terms, 
\beq
D_{11}(p)= \frac{i}{p^2 -m^2 + i \varepsilon} + 2 \pi \delta(p^2-m^2) N(|p_0|)
\eeq
the first one being just the $T=0$ propagator and the second containing
all the statistical information. The cut-off is then introduced in the 
thermal sector only,  by modifying the Bose-Einstein distribution function 
as
\beq
N(k_0) \rightarrow N_\L(k_0) = N(k_0) \theta(|\vec{k}|-\L)\;.
\eeq
Now, as $\L \rightarrow \infty $ 
the modified  propagator reduces to that of the  $T=0$ theory, 
and the generating functional gives the Green functions of the $T=0$ 
renormalized QFT ({\it i.e.} with all the $T=0$ quantum fluctuations integrated
out). By lowering $\L$, the thermal modes of momenta $|\vec{k}| > \L$
are progressively integrated out. Eventually, in the $\L \rightarrow 0$ limit, 
the $T\neq 0$ theory in equilibrium is reached. Moreover, since $N(|p_0|)$ 
comes along with a delta function forcing the momentum on the mass shell, our 
cut-off procedure does not break gauge invariance \cite{DP2}.

\section{Computation of $\gamma(T)$}
The flow equations relevant to the computation of the plasmon damping rate 
are schematically reproduced in Fig. 1. The RHS's are formally one-loop 
expression, but they contain {\it exact} and $\L$-dependent propagators and 
vertices, represented by a black dot. 
The empty dot represents the kernel of the evolution equation, which 
substitutes the full propagator in the corresponding leg. It
is given by
\[
 K_{\L,ij}(k) = - \rho_\L(k)\varepsilon (k_0) \L \delta(|\vec{k}|-\L) 
N(|k_0|)  B_{ij}\,
\]
where $\rho_\L(k)$ is the {\it full}, $\L$-dependent, spectral function, 
$\varepsilon(x)=
\theta(x)-\theta(-x)$, and $ B_{ij}=1$ with $i,j =1,2$ the thermal indices.

We need to compute the real and imaginary parts of the (1-1) component of 
the self-energy
\beq
\Sigma_{11}(\omega\pm i \varepsilon, \vec{k};\L) =
\Sigma^R_{11}(\omega, \vec{k};\L)
\pm i \Sigma^I_{11}(\omega, \vec{k};\L)
\eeq
where the real part is the same as that of the self-energy appearing in the 
propagator, 
$\Sigma^R_{11}(k;\L) =  \Pi_R(k;\L)$ and
the imaginary part is linked to $\Pi_I$ in (\ref{gammadef}) by 
$\Pi_I = \Sigma^I_{11}/(1 + 2 N)$ \cite{LvW}.

\begin{figure}
\centerline{\epsfxsize=250pt \epsfbox{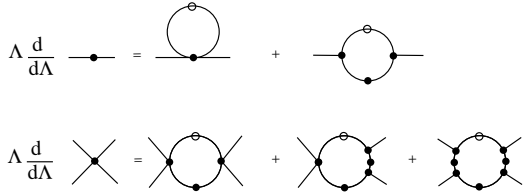}}
\caption{Schematic representation of the evolution equations for the 
two--and four-- point functions.}
\end{figure}

In order to determine the $\L$-dependent four-point function we must consider 
the system containing also the second equation in Fig.1. Since the theory has 
a -- possibly spontaneously broken -- $Z_2$ symmetry, the trilinear couplings 
are not arbitrary and will be determined from the mass and the quartic 
coupling as we will indicate below. All n-point vertices with n$>$4 will be 
neglected. 

The initial conditions for the evolution equations are given at a scale 
$\L =\L_0\gg T$ (due to the exponential damping in the Bose-Einstein function,
$\L\gta 10 \,T$ will be enough). Here, the effective action of the theory
is approximated by
\[
\Gamma_{\L_0}(\Phi) = \int d^4x \left[ \frac{1}{2} (\partial \Phi)^2 - 
\frac{1}{2} \mu^2_{\L_0} \Phi^2 - \frac{\lambda_{\L_0}}{4!} \Phi^4\right]\;,
\]
where $\mu^2_{\L_0}$ and $\lambda_{\L_0}$ are the renormalized parameters of 
the $T=0$ theory.
We are interested to the case in which the $Z_2$-symmetry is broken at 
$T=0$, so we will take $\mu^2_{\L_0} < 0$. 

Then, we make the following approximations to the full propagator and 
vertices appearing on the RHS of the evolution equations:\\
${\it i)}$ the self-energy {\it in the propagator}
is approximated by  a running 
mass, 
$\Pi(k;\L) \simeq m_\L^2$ given by 
\[ m_L^2 = \mu^2_\L \;\,({\rm if}\; \mu^2_\L>0)\;,\;\;\;\; 
- 2\mu^2_\L \;\,({\rm if} \;\mu^2_\L<0)
\]\\
${\it ii)}$ the four-point function is approximated as 
\[
\Gamma^{(4)}_\L(p_i) \simeq - \lambda_\L - i \eta_\L(p_i)\;,
\]
${\it iii)}$ the three-point function is given by
\[
\Gamma^{(3)}_\L(p_i) \simeq  0\;\,({\rm if}\; \mu^2_\L>0)\;,\;\;\;\; 
\sqrt{- 6 \lambda_\L \mu^2_\L}  \;\,({\rm if} \;\mu^2_\L<0)\,.
\]
Moreover, all vertices with at least one of the thermal indices different from
1 will be neglected.

Some comments are in order at this point. 
In perturbation theory, the on-shell imaginary part is given by the two-loop
setting sun diagram. The evolution equations of Fig.1 contain instead
only one-loop integrals, so how can an imaginary part emerge? The crucial point
here is that the momentum dependence of the imaginary part of the four-point
function has to be taken into account (see point $ii)$). 
When the latter is inserted into the
upper equation, $\Pi_I$ is generated. 

As long as the $\L$-dependent action is in the broken phase, trilinear 
couplings will be present. 
By neglecting the imaginary part of the self-energy in the propagators on the 
RHS (point {\it i)} above), we loose the contribution to $\Pi_I$ obtained by 
cutting the second diagram in the RHS of the upper equation of Fig.1. 
However, if we 
restrict ourselves to temperatures $T \ge T_C$, the trilinear couplings will be
different form zero only in a limited range of $\L$. They will not contribute
in the IR,  where the dominant contributions to the on-shell
imaginary part emerge (see Fig. 3).  
The error induced by this approximation can be estimated 
noticing that
the contribution we are neglecting is of the same nature as the one obtained 
by inserting the last diagram of the second equation in the equation for the 
two-point function. The latter is taken into account and its effect on the 
full imaginary part is of the order of a few percent.

We have now a system of four evolution equations for 
$\Pi_R(k;\L)\simeq m_\L^2$, $\Sigma^I_{11}(k;\L)$, $\lambda_\L$, and $\eta_\L(p_i)$,
with initial conditions $m_{\L_0}^2= - 2 \mu_{\L_0}^2$, $\lambda_{\L_0}$, and
$\Sigma^I_{11}(k;\L_0)=\eta_{\L_0}(p_i)=0$, respectively. 
We give, for simplicity, the explicit form of the evolution equations in the symmetric phase only,
where  the trilinear coupling vanishes;
\beqra
&\ds \L \frac{\partial\:\:}{\partial \L} &\ds m^2_\L  =
-\frac{\L^3}{4 \pi^2} \frac{N(\omega_\L)}{\omega_\L} \lambda_\L \nonumber\\
&\ds \L\frac{\partial\:\:}{\partial \L}&\ds  \lambda_\L  =  -3 \frac{\L^3}{4 \pi^2}
\frac{d\:\:}{dm_\L^2} \left(\frac{N(\omega_\L)}{\omega_\L}\right) \lambda_\L^2 \nonumber\\
&\ds \L\frac{\partial\:\:}{\partial \L}& \ds \Sigma^I_{11}(k;\L)  =
-\frac{\L^3}{4 \pi^2} \frac{N(\omega_\L)}{\omega_\L} 
\eta_\L (k,-k,q_\L,-q_\L) \nonumber \\
&\ds \L\frac{\partial\:\:}{\partial \L} & \ds \eta_\L (k,-k,q_\L,-q_\L)   =   
-\frac{\lambda_\L^2}{2} (C_\L(k_0-\omega_\L,\vec{k}) +\nonumber\\
&& \:\:\;\;\;\;\;\;\:\;\:\;\;\;\; \;\;\;\;  
\;\;\;\;\;\;\;\;\;\;\;\;\;\;\;\;\;\;\;\;\;\;\;\;\;
C_\L(k_0+\omega_\L,\vec{k}) )\nonumber
\eeqra
where $\omega_\L = (\L^2+m_\L^2)^{1/2}$, $k=(k_0,\vec{0})$, 
$q_\L=(\omega_\L, |\vec{q}|=\L)$,
and 
\[
C_\L(q_0,\vec{q}) = \frac{1}{8 \pi} \frac{N(\omega_\L)}{\omega_\L} \frac{\L}{|\vec{q}|} \left(
J(q_0)+J(-q_0)\right),
\]
with
\beqra
\ds J(q_0)& =& \ds 1 + 2 \;\theta((\L^2+q_0^2-2 \omega_\L q_0)^{1/2} -\L) 
\nonumber \\
&& \;\;\;\;\;{\mathrm if} \:\:\:\:
\left|\frac{2 \omega_\L q_0 -q_0^2 +|\vec{q}|^2}{2 |\vec{q}| \L} \right| \leq 1 \; , \nonumber \\
& = & 0 \;\;\;\;{\mathrm otherwise}. \nonumber
\eeqra
Notice that  the subsystem for $m_\L^2$ and $\lambda_\L$ is closed
and can be integrated separately.

We then  proceed as follows. Fixing the 
temperature, we first integrate the subsystem for  $m_\L^2$ and $\lambda_\L$ 
down to $\L=0$ in order to find the plasmon mass, $m^2_{\L=0}$. Then we 
fix the external momentum of $\Pi_I$ on this mass-shell, $k=(m^2_{\L=0}, 
\vec{0})$ and integrate the full system from $\L= \L_0$ down to $\L=0$.

\begin{figure}
\centerline{\epsfxsize=250pt\epsfbox{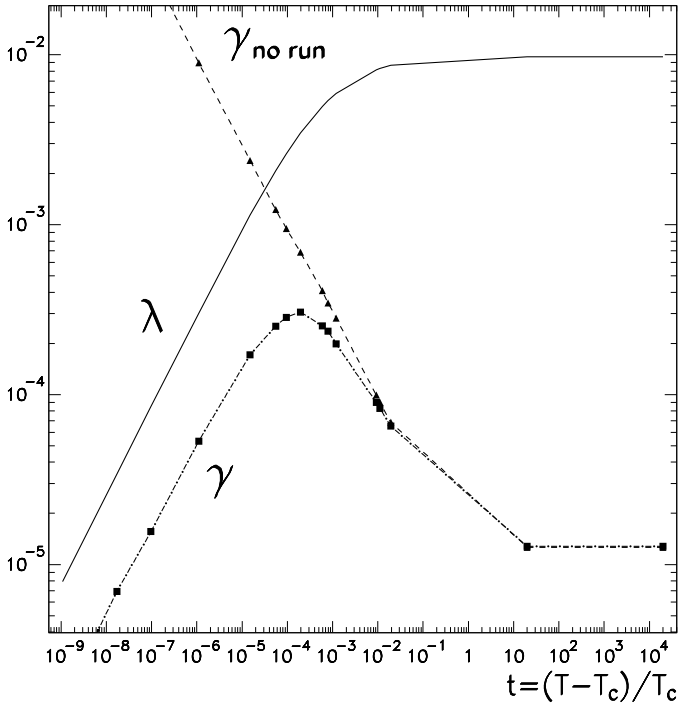}}
\caption{Temperature dependence of the coupling constant (solid line), and of 
the damping rate with the effect of the running of $\lambda$ included 
(dash-dotted) and excluded (dashed). The values for $\gamma$ have been 
multiplied by a factor of $10$.}
\end{figure}

In Fig.2 we plot the results for the damping rate $\gamma$ and the coupling 
constant at $\L=0$, as a function of the temperature \cite{Pi}. 
The dashed line has been 
obtained by keeping the coupling 
constant fixed ($\L$-independent) to its $T=0$ value ($\lambda = 10^{-2}$), 
and reproduces the divergent behavior found
in perturbation theory (eq. (\ref{gammacl})). The crucial effect of the running
of the coupling constant is seen in the behavior of the dot-dashed line. 
For temperatures close enough 
to $T_C$, the coupling constant (solid line in Fig.2) 
is dramatically renormalized and it decreases
as
\[ \lambda_{\L=0}(T) \sim t^\nu \]
where $t\equiv (T-T_C)/T_C$ and we find $\nu \simeq 0.53$ \footnote{More  
accurate results for the critical exponents from the TRG have been obtained
recently in ref. \cite{B}.}. 
The mass also vanishes with the same 
critical index. 
The decreasing of $\lambda$ drives $\gamma$ to zero, but with a different 
scaling law,
\[
\gamma_{\L=0}(T) \sim t^\nu \log t\;.
\]
The above expression can be understood noticing that the two-loop contribution
to $\Pi_I$, computed at vanishing $\omega=m(T)$,  goes as 
$\lambda^2 \log m(T)$. The RG result replaces $\lambda$ with the 
renormalized coupling, then from (\ref{gammadef}) we have 
\[\gamma \sim \frac{\lambda_{\L=0}^2}{m_{\L=0}} \log m_{\L=0} \sim 
t^\nu \log t\;.
\]

Taking couplings bigger than the one used in this letter ($\lambda =10^{-2}$), 
the deviation from the perturbative regime starts to be effective farther 
from $T_C$. Defining an effective temperature as
$\lambda_{\L=0}(T)/\lambda_{\L_0} \le 1/2$ for $T_C < T \le T_eff$ we find 
that $t_{eff}$ scales roughly 
as $t_{eff} \sim \lambda_{\L_0}$.


In Fig.3 we plot the running of $\lambda_\L(T)$ and 
$\gamma_\L(T)$ for two different values of the temperature. When $T \gg T_C$ 
most of the running takes place for 
$\L \gta \L_{soft}$, 
so it is safe to stop the
running at this scale neglecting the effect of soft loops.
When $T \rightarrow T_C$ this is not possible any longer, since most of 
the running of $\lambda$ and $\gamma$ takes place for $\L\lta\L_{soft}$.
Thus, the vanishing of the mass gap forces us to take soft  thermal momenta
into account. 


\begin{figure}
\centerline{\epsfxsize=250pt\epsfbox{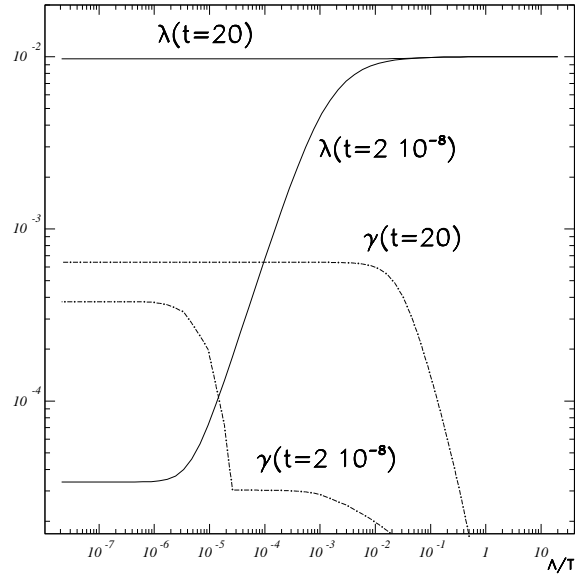}}
\caption{$\L$-running of the coupling constant $\lambda$ and of the 
damping rate $\gamma$ for two different values of $t=(T-T_C)/T_C$. The values 
of $\gamma$ (dot-dashed lines) have been multiplied by 500.}
\end{figure}

\section{Conclusion and Consequences}

The world of dynamical critical phenomena in relativistic QFT's at $T\neq 0$
is still largely unexplored, the reason being that the two main tools employed
in the static case, {\it i.e} resummed perturbation theory and lattice
simulation, are of no use in this context. I think that the TRG is on a much
better shape, and can be properly employed to
study time dependent correlation functions, relaxation rates, and transport
coefficients at {\it any} temperature, from the critical one to much higher 
(or lower) ones. 

In the case of the plasmon damping rate discussed here, the
expected critical slowing down of long wavelength
fluctuations has been obtained from first principles.
The dominant effect has been shown to be the thermal renormalization of the 
coupling constant, which turns the divergent behavior of the plasmon damping
rate into a vanishing one. 
In the future this work will be extended to non-zero spatial momentum, and to 
higher orders of approximation of the flow equations in order to check the 
scaling relations between dynamical critical exponents in a non-trivial way 
\cite{HH}.

Before concluding, let me remark two interesting physical consequences of this 
result. 

\begin{figure}
\centerline{\epsfxsize=250pt\epsfbox{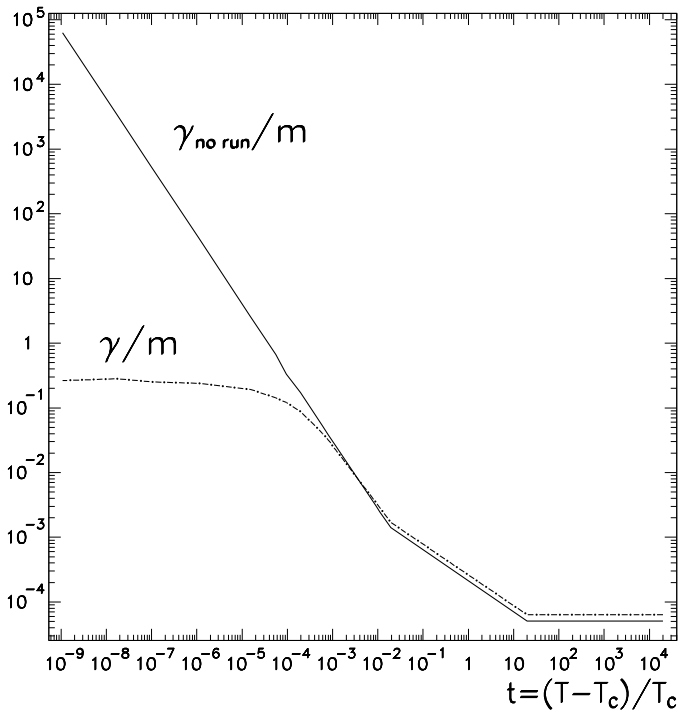}}
\caption{The ratio $\gamma/m$ vs. $T$ as obtained from TRG (dot-dashed line) 
and without including the running of $\lambda$ (solid line)} 
\end{figure}

The first consequence has to do with the notion of quasiparticle. In order for
this concept to make sense, a narrow resonance is needed, a good
criteria being $\gamma(T)/m(T) \lta 1$. In perturbation theory this
quantity diverges as $t^{-\nu}$, and one would conclude that it is not
possible to talk about quasiparticles around the critical point. 
In particular, no kinetic description close to the phase transition could
be conceivable. The TRG 
result shows that it is not necessarily the case, as we see in Fig. 4. 
Now $\gamma/m$ diverges only as $1/\log t$, and it may be  less then unity 
very much closer to the critical point.
Indeed, for $\lambda = 10^{-2}$, the ratio $\gamma/m(T)$ becomes 
larger than unity for $t\lta 10^{-5}$ in perturbation theory, whereas the 
RG result is still $\gamma/m(T) \simeq 0.3$ at $t\simeq 10^{-9}$.

The second consequence concerns second order 
phase transitions in the early universe. 
In a cosmological setting, the increasing lifetime of the fluctuations 
of the order parameter may modify the dynamics of second order -- or
weakly first order  -- phase transitions. Indeed, as the critical 
temperature is approached, the expansion rate of the universe exceeds the
thermalization rate of long wavelength fluctuations, and the latter freeze 
out. There will be a range of temperatures around $T_C$
in which the equilibrium expressions for the effective potential or the
fluctuation rates cannot be employed. In brief, second order phase transitions
in the early universe take place {\it out} of equilibrium. 

At the electroweak scale this has no practical consequences. Due to the
smallness of the expansion rate at that epoch, the temperature interval in
which there is departure from equilibrium is too short to have any 
effect. 

On the other hand, at higher temperatures, typically for $T \gta 10^{10}
$ GeV, the effect may be very strong and must be taken into account. 
This would lead to a scenario for the formation
of topological defects similar to that proposed by Zurek in \cite{Z}, and 
discussed by J. Rivers at this conference. In this scenario, 
the initial length scale of
the network of topological defects is given by the correlation length at the 
freeze-out, $\xi_{f.o.}$, whose determination requires in turn the knowledge
of the temperature dependence of $\gamma(T)$, computed here.


\begin{references}
\bibitem{DP1} M. D'Attanasio and M. Pietroni, Nucl. Phys. B472 (1996) 
711.
\bibitem{W} A. Weldon, Phys. Rev. D28 (1983) 2007.
\bibitem{BLL} D. Boyanowsky, I.D. Lawrie and D.S.  Lee, 
Phys. Rev. D54 (1996) 4013.
\bibitem{P} R.R. Parwani, Phys. Rev D45 (1992) 4695; {\it ibid.} D48
(1993) 5965 (E).
\bibitem{HW} E. Wang and U. Heinz, Phys. Rev. D53 (1996) 899
\bibitem{ASBJ1} G. Aarts and J. Smit, Pys. Lett. B393 (1997) 395; Nucl. 
Phys. B511 (1998) 451;
W. Buchm\"uller and A. Jacov\'ac, Phys. Lett. B407 (1997)
39.
\bibitem{HTL} M. Gleiser and R.O. Ramos, PRD50 (1994), 2441; C. Greiner and 
B. M\"uller, Phys. Rev. D55 (1997) 1026; P.Arnold, D. Son, and L. Yaffe,
Phys. Rev. D55 (1997) 6264; E. Iancu,  hep-ph/9710543; D. B\"odeker, 
hep-ph/9801430. 
\bibitem{BJ2} W. Buchm\"uller and A. Jacov\'ac, hep-th /9712093.
\bibitem{LK} N. Goldenfeld, {\it Lectures on Phase 
Transitions and the 
Renormalization Group}, Frontiers in Physics, Vol. 85, 
(Addison-Wesley, 1992).
\bibitem{HH} P.C. Hohenberg and B.I Halperin, Rev. Mod. Phys. 49 
(1977) 435;
J. Zinn-Justin, {\it Quantum Field Theory and Critical Phenomena}, 
International Series of Monographs in Physics, Chapter 35, (Oxford University
Press, 1996)
\bibitem{E} P.Elmfors, Z. Phys. C56 (1992) 601.
\bibitem{TW} N. Tetradis and C. Wetterich, Nucl. Phys. B398 (1993) 659.
\bibitem{Wi} K.G. Wilson, Phys. Rev. B4 (1971) 3174, 3148;
K.G. Wilson and J.G. Kogut, Phys. Rep. 12 (1974) 75.
\bibitem{Po} J. Polchinski, Nucl. Phys. B231 (1984) 269.
\bibitem{LAS} D. Litim, these proceedings; S.B. Liao and M. Strickland, 
hep-th/9803173, and  these proceedings.
\bibitem{DP2} M. D'Attanasio and M. Pietroni, Nucl. Phys. B498 (1997) 443.
\bibitem{LvW} N.P. Landsman and Ch.G. van Weert, Phys. Rep. 145 (1987) 141.
\bibitem{Pi} M. Pietroni, hep-ph/9804351, to appear on Phys. Rev. Lett.
\bibitem{B} B. Bergerhoff, hep-ph/9805493;  B. Bergerhoff and J. Reingruber, 
hep-ph/9809251.
\bibitem{Z} W.H. Zurek Phys. Rep. 276 (1996) 177.

\end{references}
\end{document}